\documentclass[unnumsec,webpdf,modern,medium,namedate]{oup-authoring-template}
\onecolumn

\usepackage{url}
\usepackage{lineno}
\usepackage{threeparttable}
\usepackage{wrapfig}
\usepackage{lscape}
\usepackage{rotating}

\usepackage{amsmath}
\usepackage{amssymb}
\usepackage{amsthm}
\usepackage{bm}
\usepackage{accents}

\usepackage{graphicx}
\usepackage{psfrag,epsf} 
\usepackage{tabularx}
\usepackage{booktabs}
\usepackage{multirow}

\usepackage{algorithm}
\usepackage{algpseudocode}  

\usepackage{natbib}

\usepackage{hyperref}

\begin{document}

\journaltitle{Statistical Analysis and Data Mining: An ASA Data Science Journal}
\copyrightyear{2025}
\firstpage{1}


\title[Modeling Spatial Heterogeneity based on Regularized Adaptive-minimum-spanning Tree]{FLAT: Fused Lasso with Adaptive-minimum-spanning Tree with Applications on Thermohaline Circulation} 

\author[1]{Cuiwen Che}
\author[1]{Yifan Chen}
\author[2,$\ast$]{Zhaoyu Xing \ORCID{0009-0002-4344-8684}}
\author[1]{Wei Zhong}

\authormark{Che et al.}

\address[1]{\orgdiv{Department of Statistics and Data Science}, \orgname{Xiamen University}, \orgaddress{\street{422 Siming South Road}, \postcode{361005}, \state{Fujian}, \country{China}}}
\address[2]{\orgdiv{Department of Applied and Computational Mathematics and Statistics}, \orgname{University of Notre Dame}, \orgaddress{\street{Holy Cross Dr}, \postcode{46556}, \state{Indiana}, \country{U.S.A.}}}

\corresp[$\ast$]{Corresponding author. \href{zxing@nd.edu}{zxing@nd.edu}}

\abstract{
This article introduces a new methodology model both discrete and continuous spatial heterogeneity simultaneously with an application in detection of hyper-plain in thermohaline circulation. To enable the data-driven detection of spatial boundaries with heterogeneity, we constructs an adaptive minimum spanning tree guided by both spatial proximity and coefficient dissimilarity, and combines both a spatial fused regularization and LASSO-type regularization to estimate the spatial coefficients under the framework of spatial regression. Numerical simulations demonstrate the effectiveness of proposed method in both estimation and heterogeneity detection. The usefulness of the approach is further illustrated via an analysis of oceanic data that provides new empirical finds about Atlantic with detected surfaces in temperature-salinity relationship.}





\maketitle

\section{Introduction}  
\label{S1}  

Spatial heterogeneity commonly exist in multiple disinclines and plays a pivotal role in oceanic and earn system. The analyses of spatial heterogeneity provide critical information about the spatially varying relationships, such as soil moisture-temperature relationships across space \citep{zhang2024learning} and sudden shifts in rainfall regimes along mountain ridges \citep{luo2021bayesian}. In ocean science, the temperature-
salinity (T-S) relationship in thermohaline circulation are also widely studied in oceanic studies \citep{
talley2011descriptive,talley2013closure,li2021thermohaline}. 

One of the key challenges to model the spatial heterogeneity is to exhibits a dual nature of spatial non-stationarity of a complex system, with both smooth variations within specific regions and discrete transitions at their boundaries \citep{fotheringham2017multiscale}. They are also known as continuous spatial heterogeneity \citep{bailey1995interactive} and discrete heterogeneity \citep{ferrari2001temperature, anselin2010thirty} respectively. 


To accurately model spatial heterogeneity, several models were developed \citep{lesage2009introduction, ward2018spatial} under the framework of spatial regression, where the local spatial dependencies are characterized by the a set of spatial coefficients \citep{brunsdon1996geographically}, such as the geographically weighted regression (GWR) model \citep{brunsdon1996geographically}, the spatial lag model \citep{lam2020estimation}, and the spatial Durbin model \citep{anselin2013spatial}. However, these models can only capture the continuous spatial heterogeneity without extra design to detect potential spatial boundaries, such as the inverse T-S surface in thermohaline circulation. Although there are empirical studies that specially designed two-step procedures to capture the discrete spatial heterogeneity \citep{chen2024heterogeneity, duque2011p} with clustering-based segmentation techniques \citep{raudenbush2002hierarchical, van2019modernizing}, they are not general data-driven methods for modeling spatial heterogeneity. 

However, the detection of boundaries with sharp changes in spatial coefficients are important in many applications. As the T-S relationships is the primary driver of thermohaline circulation that plays a crucial role in regulating Earth's climate, shaping marine ecosystems, and influencing sea-level rise \citep{rahmstorf2003thermohaline, velasco2021synergistic, lin2023scale}. Other examples of the discrete spatial heterogeneity can be widely find in studies of spatial modeling, such as economic boundaries \citep{GelfandJASA} and areal wambling \citep{kudela2015application, luo2021bayesian}. Real-world geographic spaces often exhibit both continuous and discrete spatial heterogeneity, but there are few general data-driven methods proposed for general modeling local smoothness while also identifying spatial boundaries adaptively. 


To generally model the spatial heterogeneity with the dual nature, we propose a new approach, namely Fused Lasso with an Adaptive-minimum-spanning Tree (FLAT), which can effectively estimate the spatial coefficients in spatial regression model while capturing the discrete spatial heterogeneity by detecting the spatial boundaries simultaneously. Specifically, FLAT constructs fused penalization terms guided by minimum spanning tree structures that incorporate both spatial proximity and coefficient similarity, and introduces a structured adaptive regularization to model both continuous and discrete spatial heterogeneity in a unified optimization procedure. In addition, FLAT utilizes a density-based spatial clustering method to further detect the non-convex spatial regions with heterogeneity in applications. The comprehensive simulations demonstrate the dominant performance of FLAT in both parameter estimation and the detection of boundaries. A real-world oceanic application to model thermohaline circulation with new empirical finds demonstrates the effectiveness of FLAT in practice.

The rest of the article is organized as follows.  Section \ref{S2} introduces the spatial regression modal and the methodology of FLAT. Section \ref{S3} presents the simulation studies showing the effectiveness of FLAT. A real application of FLAT on thermohaline circulation is shown in Section \ref{S4} with analysis of new finds of North Atlantic Ocean. Section \ref{S5} concludes this article with a discussion. The technical details of both the algorithm and the real data analysis are given in Appendix.

\section{Fused Lasso Regression With Adaptive Minimum Spanning Tree}   
\label{S2}       

In this paper, we focus on modeling the complex relationships and heterogeneity in spatial data $\{ \bm{x}(s_i), y(s_i) \}$ for $i\in[n]=\{1,\cdots,n\}$, where $\bm{x}(s_i)=[x_1(s_i),\cdots,x_p(s_i)]^\top$ is a $p$-dimensional vector of spatial features at the spatial location $s_i$, and $y(s_i)$ is the corresponding response. Then we consider a spatial regression model \citep{ward2018spatial} with spatially varying coefficient as below: 
\begin{equation} 
    y(s_i) = \boldsymbol{\beta}^{\top}(s_i) \bm{x}(s_i) + \epsilon(s_i), 
\end{equation} 
where $\boldsymbol{\beta}(s_i) = [ \beta_1(s_i) \cdots, \beta_p(s_i) ]^\top$ is the heterogeneous coefficient vector at spatial location $s_i$, $\epsilon(s_i)$ is independent random error term with mean $0$ and variance $\sigma^2$. As introduced in Section \ref{S1}, discrete spatial heterogeneity is widely considered in empirical analysis of spatial modeling, which is captured by the spatial clusters among $\boldsymbol{\beta}(s_i)$ in spatial regression models. 

To fully capture the heterogeneity in spatial data and detect the potential clusters with spatial boundaries, we consider to capture the spatial dependency between coefficients utilizing the minimum spanning tree, and then construct the fused regularization for estimation.

To generate the minimum spanning tree $\mathbb{E}$ considering both spatial distance and the spatial heterogeneity, we firstly generated the spatial linkage structure $\mathbb{E}_0$ capturing the pure spatial distance spatial based on spatial distances $d(s_i, s_j) = \| s_i  -  s_j\|_2$. Then we obtain the estimator $\hat{\bm{\beta}}^0(s_i), i\in[n]$ for parameter $\beta(s_i)$ at spatial location $s_i$ considering pure spatial distance by optimizing
\begin{equation}
\mathop{\arg\min}\limits_{\boldsymbol{\beta}}\frac{1}{n}\sum\limits_{i=1}^{n}\left(y(s_i)-\boldsymbol{\beta}^\top(s_i)\bm{x}(s_i)\right)^2+\lambda\sum\limits_{k=1}^{p}\sum\limits_{(s_i,s_j)\in\mathbb{E}_0}|\beta_k(s_i)-\beta_k(s_j)|.
\end{equation} 
To fully extract the information of spatial heterogeneity, we generate the full graph $\mathbb{E}_1$ among spatial locations based on the distance $\tilde{d}(s_i, s_j)$ defined as the $L_2$-norm of the difference in parameters between spatial locations, that is
\begin{equation}
\label{eq:beta_distance}
      \tilde{d}(s_i, s_j) = \|\hat{\boldsymbol{\beta}}^0(s_i) - \hat{\boldsymbol{\beta}}^0(s_j)\|_2.
\end{equation}
Then the minimum spanning tree $\mathbb{E}$ is obtained from $\mathbb{E}_1$ via Prim's algorithm. Based on the definition of minimum spanning tree, the spatial linkage structure $\mathbb{E}$ will contain $n-1$ links.

To be consistent with the classic literature \citep{zhang2024learning, liSangJASA2019spatial}, we construct the connectivity matrix $\bm{H} \in \mathbb{R}^{(n-1) \times n}$ from $\mathbb{E}$, where element $\bm{H}_{l,i}$ of $\bm{H}$ on $l$-th row and $i$-th column os defined based on the connection status of vertices $i\in[n]$ along the edges $l\in[n-1]$. If the $l$-th potential edge, which starts from of spatial location $s_i$ to $s_j$, exists in the spatial linkage structure $\mathbb{E}$, then we set $\bm{H}_{l,i} = 1$, $\bm{H}_{l,j} = -1$, and set all the other elements in $l$-th row of matrix $\bm{H}$ as $0$. To capture discrete spatial heterogeneity in the spatial regression, we utilize the representation of minimum spanning tree above and consider the following regularization 
\begin{equation}
\label{Hbeta1}
\|\bm{H} \boldsymbol{\beta}_k \|_1 = \sum_{(s_i, s_j) \in \mathbb{E}} |\bm{\beta}_k(s_i) - \bm{\beta}_k(s_j)|,
\end{equation}
for $k$-th dimension of the spatial coefficient $\bm{\beta}$, where $\{ \boldsymbol{\beta}_k = ( \beta_k(s_1), ..., \beta_k(s_n)\}^{\top}, \quad k = 1, ..., p$ represent the coefficients of the $k$-th explanatory variable among all spatial locations. Thus, $\|\bm{H} \boldsymbol{\beta}_k \|_1$ represents regularization based on the adaptive minimum spanning tree $\mathbb{E}$ in terms of the $k$-th spatial features. 


To further includes the continuous spatial heterogeneity with the regularization defined in \eqref{Hbeta1}, we further denote $\pi_{(s_i, s_j)} = 1 / \tilde{d}(s_i, s_j)^\gamma $ as the adaptive weights between any two spatial \((s_i, s_j)\) for all $(s_i, s_j) \in \mathbb{E}$ and $0$ otherwise. Although $\gamma$ can be any positive constant to control the adaptiveness of regularization, we follow the recommendation of \citet{zou2006adaptive} and set $\gamma=1$ in both simulations in Section \ref{S3} and real-data analysis in Section \ref{S4}. 
Then we propose the objective function of FLAT as  
\begin{equation}
\label{FLAT_original}
    \hat{\boldsymbol{\beta}}_{\mathrm{FLAT}} = \arg\min_{\beta} \frac{1}{n} \sum_{i=1}^{n} \left\{ y(s_i) - \boldsymbol{\beta}^\top(s_i) \bm{x}(s_i) \right\}^2 + \lambda_1 \sum_{k=1}^{p} \| \bm{\pi} \bm{H} \bm{\beta}_k \|_1 + \lambda_2 \sum_{k=1}^{p} \| \bm{\beta}_k \|_1,
\end{equation}
where $\lambda_1$ and $\lambda_2$ are tuning parameters controlling the strength of penalization to capture the discrete spatial heterogeneity and the sparsity in spatial features. The adaptive weights $\bm{\pi}$ further enhance the detection of spatial heterogeneity while reducing the estimation error for the spatial locations with homogeneity. With the specially designed regularization based on the adaptive-minimum-spanning
tree, we can effectively detect the discrete spatial heterogeneity and select the spatial features simultaneously while estimate the continuous spatial coefficients by optimizing \eqref{FLAT_original}. 
The details of the derivation and the algorithms are given in Section \ref{FLATdetailsalgo} of Appendix, and the R tools for FLAT are publicly available\footnote{\url{https://github.com/ZhaoyuXingStat/FLAT}}. We next investigate the numerical performance of $    \hat{\boldsymbol{\beta}}_{\mathrm{FLAT}}$ in the following sections.

\section{Simulation}  
\label{S3}


\subsection{Simulation settings}
\label{S3.1}
To fully characterize marine conditions where the response variables of regression models often exhibit distinct spatial structures in geoscientific research, we consider two-dimensional spatial data in the simulations, which is commonly considered in marine across longitude and latitude coordinates or longitude and depth. Without losing generality, we assume a $[0,1]\times[0,1]$ spatial domain with $n=1,000$ randomly distributed observation points $s_1,\dots,s_n\in\mathbb{R}^2$.    

In all settings of our simulations, the responses at spatial locations $\{s_1,\dots,s_n\}$ are generated based on spatial regression model 
\begin{equation}
\label{spatialRM}
y(s_i)=\beta_1(s_i)x_1(s_i)+\beta_2(s_i)x_2(s_i)+\beta_3(s_i) + \epsilon(s_i),\quad i=1,\dots,n,
\end{equation}
where $\epsilon(s_i)$ represent the independent normal error terms with zero mean and standard deviation $\sigma=0.1$. We use different approaches to estimate the parameters $\{\beta_1(s_i),\beta_2(s_i),\beta_3(s_i)\}_{i\in[n]}$ based on the generated spatial data $\{ \bm{x}(s_i), y(s_i)\}_{i\in[n]}$ and compare their performance in capturing both continuous spatial heterogeneity and discrete spatial heterogeneity.

The spatial dependence among spatial covariates is specially designed with different covariance structure. In all simulations, two independent Gaussian stochastic processes $\{z_1(s_i)\}_{i=1}^{n}$ and $\{z_2(s_i)\}_{i=1}^{n}$ are generated with zero mean and designed covariances $\text{cov} (z_k(s_i),z_k(s_j) )=\exp(-\|s_i-s_j\|_2 / \phi ),\ k=1,2.$

In the generating process, parameter $\phi$ controls the spatial heterogeneity.  Then the spatial features $x_1(s_i)$ and $x_2(s_i)$ are generated from $z_1(s_i)$ and $z_2(s_i)$ by linear transformation. Specifically, we set $x_1(s_i)=z_1(s_i)$ and $x_2(s_i)=rz_1(s_i)+\sqrt{1-r^2}z_2(s_i)$, 
where the parameter $r$ controls the level of collinearity. We set $r=0.75$ for a strong correlation design and generate five datasets with $\phi=0.2,0.4,0.6, 0.8,1$ for different levels of spatial correlation. The true values of the parameters $\beta_1,\beta_2,\beta_3$ are set as shown in Figure \ref{fig3.1}(a)-(c) respectively, where different clustering patterns are adopted for three different parameters. 
 
To quantify the performance of estimations for dimension $k\in[p]$ and location $i\in [n]$, we focus on the three measures with $T=100$ independent replications: (1) Root Mean Squared Error (RMSE), which is defined as $\text{RMSE}_{k}^{(i)}=[\sum_{t=1}^{T}\{\hat{\beta}_k^t(s_i)-\beta_k(s_i)\}^2 /T ] ^{1/2}$ (2) Mean Absolute Error (MAE), which is defined as $\text{MAE}_{k}^{(i)}= \sum_{t=1}^{T}|\hat{\beta}_k^t(s_i) - \beta_k(s_i)|/T$; (3) Standard Deviation (SD), which is defined as $\text{SD}_{k}^{(i)}= [ \sum_{t=1}^{T}\{\hat{\beta}_k^t(s_i)-\bar{\hat{\beta}}_k(s_i)\}^2 /T  ]^{1/2}$, where $\bar{\hat{\beta}}_k(s_i)=\sum_{t=1}^{T}\hat{\beta}_k^t(s_i)/T$. Smaller values in RMSE, MAE and SD indicate more accurate and stable estimations for parameters in model \eqref{spatialRM}. As we have in total $np$ parameters to estimate, the mean values of the aforementioned measures across all spatial points $i\in[n]$ for each dimension $k\in[p]$ are computed and reported in Table \ref{tab3.1}. 

To assess the performance in detecting discrete spatial heterogeneity in each dimension $k\in[p]$, we consider four widely used measures for clustering: Rand Index (RI), Adjusted Rand Index (ARI), Silhouette Coefficient (SC) and Calinski-Harabasz Index (CHI). Higher values of RI, ARI, SC and CHI reflect a greater consistency between the detected clusters and the true group labels. Detailed definitions and further discussion of these metrics are included in \ref{DEC}.

We compare our FLAT method with two classic works: Geographically Weighted Regression (GWR) model \citep{brunsdon1996geographically} and Spatially Clustered Coefficient (SCC) model \citep{liSangJASA2019spatial}. To mitigate the impact of errors due to randomness, we conduct $T=100$ repeats for each setting and evaluate the estimating and clustering performance of different methods.

\subsection{Simulation results}
\label{S3.3}

\begin{figure}
	\centering
    \includegraphics[width=0.9\textwidth]{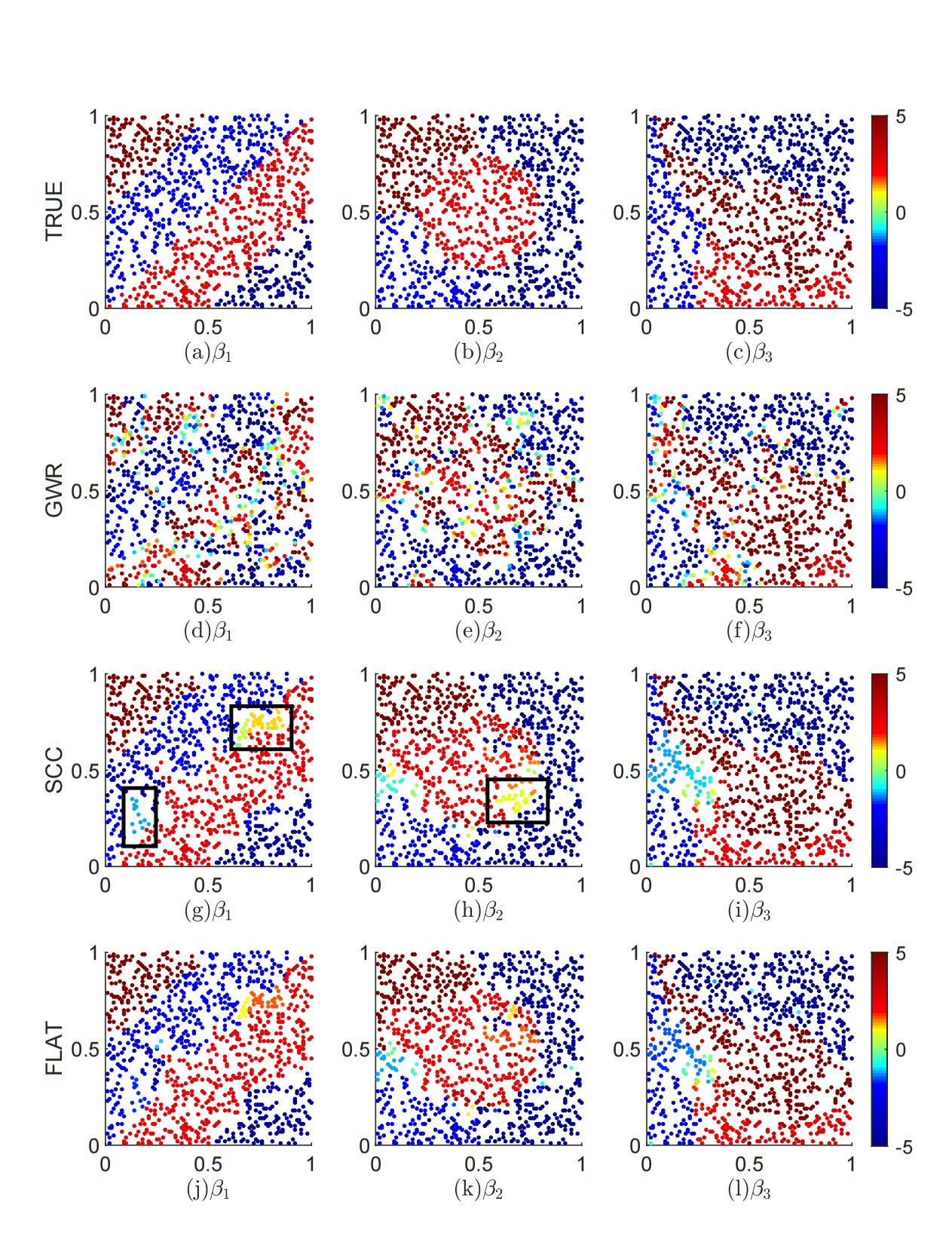}
    \caption{True spatial parameters $\beta_k$ and estimated parameters $\hat{\beta}_k$ for $k=1,2,3$ by GWR, SCC and FLAT: In each subgraph, the coordinate axes correspond to two spatial dimensions and the points represent the spatial locations $s_i$ accordingly. The dataset is generated from DGP5 and the values of parameters and estimations are indicated by color.}  
    \label{fig3.1}
\end{figure}

We firstly illustrate the finite sample performance of our method FLAT through scatter plots in Figure \ref{fig3.1} and provide a direct visual comparison of estimated coefficients by different methods. The first row with three plots (a)-(c) shows the true parameters in data-generating process, and other rows show the estimated parameters by GWR, SCC and proposed FLAT, respectively. The three columns represent the spatial coefficients $\beta_1(s_i)$, $\beta_2(s_i)$ and $\beta_3(s_i)$, and the color represents the quantities of the parameters. From the sub-figure (a)-(c), there's different layers in each scatter plot, which indicates the cluster patters in different dimension. Sub-figure (d)-(f) shows that GWR can hardly detect the heterogeneity in spatial parameters, and FLAT provides more accurate estimations in sub-figure (j)-(l) compared to GWR. While the SCC can detect part of the spatial heterogeneity, FLAT excels in boundary identifications as marked in the boxed areas of sub-figure (g) and (h). 

\begin{table}[]
    \caption{The comparison of estimation performance with different levels of spatial correlation: The mean of the different measures across all spatial points $i\in[n]$ are computed and reported for each individual parameter with different level of spatical correlation.}
    \label{tab3.1}
    \centering
    \footnotesize 
    \begin{tabular}{@{}cc rrr rrr rrr@{}}
        \toprule
        & & \multicolumn{3}{c}{GWR} & \multicolumn{3}{c}{SCC} & \multicolumn{3}{c}{FLAT} \\
        \cmidrule(lr){3-5} \cmidrule(lr){6-8} \cmidrule(lr){9-11}
        Spatial Correlation $\phi$ & Evaluation & $\beta_1$ & $\beta_2$ & $\beta_3$ & $\beta_1$ & $\beta_2$ & $\beta_3$ & $\beta_1$ & $\beta_2$ & $\beta_3$ \\
        \midrule
        \multirow{3}{*}{0.2} & RMSE & 3.0521 & 3.0344 & 3.0311 & 0.8612 & 0.9390 & 0.9015 & 0.8090 & 0.8894 & 0.8502 \\
        & MAE  & 2.3340 & 2.3247 & 2.0214 & 0.5856 & 0.6520 & 0.5698 & 0.5370 & 0.5940 & 0.5395 \\
        & SD   & 2.9521 & 2.9500 & 2.7790 & 0.7213 & 0.7721 & 0.7644 & 0.6888 & 0.7501 & 0.7151 \\
        \midrule
        \multirow{3}{*}{0.4} & RMSE & 4.2638 & 4.2502 & 4.5394 & 1.0648 & 1.1372 & 1.1835 & 0.9800 & 1.0454 & 1.0845 \\
        & MAE  & 3.2390 & 3.2121 & 2.8968 & 0.7330 & 0.8019 & 0.7448 & 0.6645 & 0.7191 & 0.6884 \\
        & SD   & 4.1770 & 4.1666 & 4.3506 & 1.4430 & 1.7088 & 1.8384 & 0.8401 & 0.8764 & 0.9165 \\
        \midrule
        \multirow{3}{*}{0.6} & RMSE & 5.1742 & 5.1628 & 5.6082 & 1.2451 & 1.3132 & 1.4219 & 1.1503 & 1.2160 & 1.3177 \\
        & MAE  & 3.9096 & 3.8829 & 3.5364 & 0.8726 & 0.9443 & 0.8997 & 0.7906 & 0.8520 & 0.8280 \\
        & SD   & 5.0925 & 5.0833 & 5.4442 & 1.0591 & 1.0781 & 1.2170 & 0.9865 & 1.0102 & 1.1237 \\
        \midrule
        \multirow{3}{*}{0.8} & RMSE & 5.9329 & 5.9215 & 6.4590 & 1.3882 & 1.4588 & 1.5988 & 1.2817 & 1.3582 & 1.4848 \\
        & MAE  & 4.4751 & 4.4486 & 4.0649 & 0.9953 & 1.0641 & 1.0234 & 0.8984 & 0.9684 & 0.9445 \\
        & SD   & 5.8530 & 5.8433 & 6.3084 & 1.1813 & 1.1950 & 1.3703 & 1.0991 & 1.1270 & 1.2681 \\
        \midrule
        \multirow{3}{*}{1} & RMSE & 6.7977 & 6.8203 & 7.5956 & 1.5690 & 1.6872 & 1.8425 & 1.4446 & 1.5665 & 1.7280 \\
        & MAE  & 5.0945 & 5.0979 & 4.7283 & 1.1371 & 1.2616 & 1.2073 & 1.0293 & 1.1459 & 1.1196 \\
        & SD   & 6.7234 & 6.7586 & 7.4337 & 1.3341 & 1.3522 & 1.5622 & 1.2338 & 1.2729 & 1.4605 \\
        \bottomrule
    \end{tabular}
\end{table}

The comparison based on the measures of estimations shows the similar conclusion in Table \ref{tab3.1}, where the mean values of the different measures across all spatial points $i\in[n]$ for each dimension $k\in[p]$ are computed and reported for each individual parameter $\beta_1$, $\beta_2$ and $\beta_3$ under the different data-generating settings. For each dimension of spatial parameter $\bm{\beta}$, FLAT achieves the dominated performance in estimation in terms of RMSE, MAE and SD compared to the classic methods GWR and SCC. Besides, FLAT shows more stable estimations with increasing local dependency from DGP1 to DGP5, while GWR can hardly handle the strong spatial correlation. 

The comparison based on the measures of clustering is shown in Table \ref{tab3.2}. As shown in the first row of Figure \ref{fig3.1}, true spatial parameters have spatial clustering patterns. We conduct the identical standard DBSCAN clustering algorithm with the estimations by GWR, SCC and FLAT and calculate the RI, ARI, SC, and CHI, which measure the similarity between the discrete spatial heterogeneity and the detected clusters of coefficients in different dimensions. Thus, more accurate clustering results will indicate better capture of discrete spatial heterogeneity. FLAT achieves better performance with higher RI, ARI, SC and CHI compared to the classic methods GWR and SCC, which indicates the effectiveness of FLAT in detecting the discrete spatial heterogeneity in practice.


\begin{table} 
		\caption{Evaluations of estimation performance based on clustering with different level of spatial correlation: RI represents the Rand index, ARI represents the adjusted Rand index, SC represents the Silhouette coefficient, and CHI represents Calinski-Harabasz index and scaled by $\times10^4$.}
        \label{tab3.2}
		\centering
        \footnotesize
\begin{tabular}{@{}ccrrrrrrrrr@{}}
\toprule
                      &                   & \multicolumn{3}{c}{GWR}           & \multicolumn{3}{c}{SCC}              & \multicolumn{3}{c}{FLAT}          \\  \cmidrule(lr){3-5} \cmidrule(lr){6-8} \cmidrule(lr){9-11}
Spatial Correlation $\phi$                   & Evaluation        & $\beta_1$ & $\beta_2$ & $\beta_3$ & $\beta_1$ & $\beta_2$ & $\beta_3$ & $\beta_1$ & $\beta_2$ & $\beta_3$ \\ \midrule
\multirow{4}{*}{0.2} & RI                & 0.6411    & 0.7446    & 0.8084    & 0.9117    & 0.9185    & 0.9346    & 0.9180    & 0.9289    & 0.9370    \\
                      & ARI               & 0.2453    & 0.2769    & 0.3576    & 0.7898    & 0.7490    & 0.7709    & 0.8051    & 0.7804    & 0.7818    \\
                      & SC                & 0.4537    & 0.4163    & 0.5163    & 0.8685    & 0.8378    & 0.8713    & 0.8791    & 0.8577    & 0.8835    \\
                      & CHI & 0.2041    & 0.5611    & 0.9952    & 2.3816    & 3.1291    & 4.3364    & 3.3122    & 4.0711    & 4.7318    \\ \midrule
\multirow{4}{*}{0.4} & RI                & 0.6220    & 0.7210    & 0.7800    & 0.8816    & 0.8942    & 0.9002    & 0.8932    & 0.9074    & 0.9107    \\
                      & ARI               & 0.1843    & 0.2173    & 0.2933    & 0.7280    & 0.6826    & 0.6753    & 0.7519    & 0.7176    & 0.7056    \\
                      & SC                & 0.3160    & 0.3331    & 0.4142    & 0.8373    & 0.8137    & 0.8432    & 0.8492    & 0.8307    & 0.8441    \\
                      & CHI  & 0.2232    & 0.5846    & 1.0047    & 1.5141    & 2.4471    & 2.9139    & 2.0688    & 3.2222    & 3.6620    \\ \midrule
\multirow{4}{*}{0.6} & RI                & 0.6298    & 0.7027    & 0.7619    & 0.8525    & 0.8732    & 0.8832    & 0.8697    & 0.8904    & 0.8959    \\
                      & ARI               & 0.1702    & 0.1853    & 0.2607    & 0.6607    & 0.6185    & 0.6296    & 0.6946    & 0.6672    & 0.6589    \\
                      & SC                & 0.2606    & 0.2720    & 0.3454    & 0.7910    & 0.7928    & 0.8217    & 0.8236    & 0.8204    & 0.8281    \\
                      & CHI  & 0.2975    & 0.6311    & 1.0801    & 1.2801    & 2.0624    & 2.4289    & 1.8807    & 2.6647    & 2.8204    \\ \midrule
\multirow{4}{*}{0.8} & RI                & 0.6206    & 0.6968    & 0.7467    & 0.8286    & 0.8559    & 0.8661    & 0.8516    & 0.8674    & 0.8799    \\
                      & ARI               & 0.1481    & 0.1681    & 0.2370    & 0.6071    & 0.5701    & 0.5871    & 0.6545    & 0.6030    & 0.6150    \\
                      & SC                & 0.2246    & 0.2144    & 0.2862    & 0.7702    & 0.7715    & 0.7931    & 0.7990    & 0.7841    & 0.8172    \\
                      & CHI  & 0.3315    & 0.7392    & 1.1285    & 1.0753    & 1.8524    & 2.0760    & 1.5322    & 2.5714    & 2.4064    \\ \midrule
\multirow{4}{*}{1} & RI                & 0.6101    & 0.6760    & 0.7176    & 0.8056    & 0.8307    & 0.8640    & 0.8227    & 0.8471    & 0.8718    \\
                      & ARI               & 0.1260    & 0.1487    & 0.2061    & 0.5650    & 0.5023    & 0.5660    & 0.5951    & 0.5386    & 0.5892    \\
                      & SC                & 0.1879    & 0.1363    & 0.2091    & 0.7494    & 0.7572    & 0.7871    & 0.7680    & 0.7754    & 0.8066    \\
                      & CHI & 0.3858    & 0.7457    & 1.3860    & 0.7449    & 1.6862    & 2.0479    & 1.0421    & 2.3999    & 2.1975    \\ \bottomrule
\end{tabular}

	\end{table}


\section{Real Data Analysis}  
\label{S4}

We apply the FLAT method to study the seawater temperature-salinity (T-S) relationship, which is a critical oceanographic topic with both scientific and practical relevance. In physical ocean dynamics, seawater density is jointly influenced by temperature (which decreases density) and salinity (which increases density). When horizontal gradients of temperature and salinity are spatially opposed, their effects on density become reinforcing \citep{flament1985evolving, rudnick1996intensive}. This phenomenon, known as Temperature-Salinity (T-S) compensation, arises from the counteracting contributions of thermal and haline gradients. Globally, density distributions are largely governed by thermal gradients due to their generally larger magnitude compared to salinity gradients \citep{reid1969sea, roden1975north}.

Ocean stratification is fundamentally tied to T-S gradients. Warmer or fresher (less dense) waters tend to overlie colder or saltier (denser) layers, exemplified by Antarctic Intermediate Water. This stratification shapes vertical mixing, nutrient fluxes, and biogeochemical cycles. To quantify spatial gradients, we adopt a numerical method called the spatial difference quotient, which measures the rate of change across discrete spatial datasets-analogous to the derivative in continuous calculus. We introduce these concepts and corresponding formulas in \ref{DRDA}. 

\subsection{Dataset and description}

We use the public oceanic data from the Copernicus Marine Environment Monitoring Service (CMEMS)\footnote{\url{https://marine.copernicus.eu}} in real data analysis and investigate the T-S relationships via FLAT. CMEMS is a major global ocean observation and forecasting initiative, initiated and coordinated by the European Union, with operational services delivered by Mercator Ocean International (France) to users worldwide. This study utilizes the gridded, daily-averaged physical oceanographic dataset ``GLOBAL\_MULTIYEAR\_001\_030” from the GLORYS12V1 product suite under CMEMS. The dataset provides vertical profiles of temperature, salinity, ocean currents, sea level, mixed layer depth, and sea ice parameters, with temporal coverage from 1993 to the present.

This study selected two datasets as the focus of our investigation. The first dataset comprises sea surface T-S data for the North Atlantic Ocean, spanning a meridional range from 0° to 60°W and a zonal range from 0° to 60°N. The second dataset includes T-S observations along a longitudinal transect at 25°W across the Atlantic Basin, extending from 60°N to 60°S latitude and encompassing 43 discrete depth levels between 0 and 3000 meters. This cross-sectional profile, referred to as an oceanographic section, represents a widely adopted approach in marine research for characterizing vertical and horizontal gradients in hydrographic properties.

\subsection{Estimation results}

We consider the spatial regression modeling between temperature and salinity as
\begin{equation}
    \mathcal{S}(s_i)=\beta_0(s_i)+\beta_1(s_i)\mathcal{T}(s_i)+\epsilon(s_i),
\end{equation}
where $\mathcal{S}$ and $\mathcal{T}$ represent the salinity and temperature of seawater, respectively. Then we optimize the FLAT objective function as below
\begin{align}
    L(\boldsymbol{\beta})= &\frac{1}{n}\sum\limits_{i=1}^{n}\left(\mathcal{S}(s_i)-\beta_1(s_i)\mathcal{T}(s_i)-\beta_0(s_i)\right)^2 \nonumber\\
    &+\lambda_1\sum\limits_{(s_i,s_j)\in\mathbb{E}}\|\beta_1(s_i)-\beta_1(s_j)\|_1+\lambda_2\sum\limits_{i=1}^{n}\|\beta_1(s_i)\|_1
\end{align}
and obtain the estimated coefficients for all the spatial locations in the marine data. The tuning and clustering procedures are the same as simulations in Section \ref{S3}. The evaluation of estimations based on RMSE and the performance of clustering are summarized in Table \ref{tab4.2}. We find that among the sea level data and ocean section profile data, FLAT exhibits the lowest RMSE, indicating that it provides the best fitting performance. Also, FLAT achieves best clustering results in terms of both SC and CHI in the marine surface data, suggesting an outperforming estimation compared to GWR and SCC. Overall, best results in both datasets demonstrates the advantage of FLAT in detecting the discrete spatial heterogeneity in T-S relationships.
\begin{table}[]
		\caption{Estimation results for T-S relationships in oceanic data}
        \label{tab4.2} 
        \centering
		\begin{tabular}{ccccccccccccc}
		\toprule
		  \multirow{2}*{Method} & & \multicolumn{5}{c}{Sea level} & & \multicolumn{5}{c}{Sea cross-section} \\ 
          & $\ \ $ & RMSE & $\ $ & SC & $\ $ & CHI($\times10^3$) & $\ \ $ & RMSE & $\ $ & SC & $\ $ & CHI($\times10^3$) \\     \hline
			GWR  & & 0.2879 & & 0.0175 & & 0.4079 & & 0.0827 & & 0.0078 & & 0.1191 \\ 
			SCC  & & 0.1989 & & 0.3293 & & 0.7108 & & 0.0546 & & 0.1093 & & 1.6032 \\ 
            FLAT & & \textbf{0.1941} & & \textbf{0.4334} & & \textbf{1.9654} & & \textbf{0.0500} & & \textbf{0.2090} & & \textbf{1.8590} \\ 
            \bottomrule
		\end{tabular} 
	\end{table} 

\subsection{Analysis of the T-S relationship at sea surface layers}

In this section, we closely show the estimation results by FLAT on sea-level data in the meridional range from 0° to 60°W and in the zonal range from 0° to 60°N and several valuable findings based on the discrete spatial heterogeneity detected. 

\begin{figure}[H]
    \centering
    \begin{minipage}{0.47\textwidth}
        \centering
        \includegraphics[width=\textwidth]{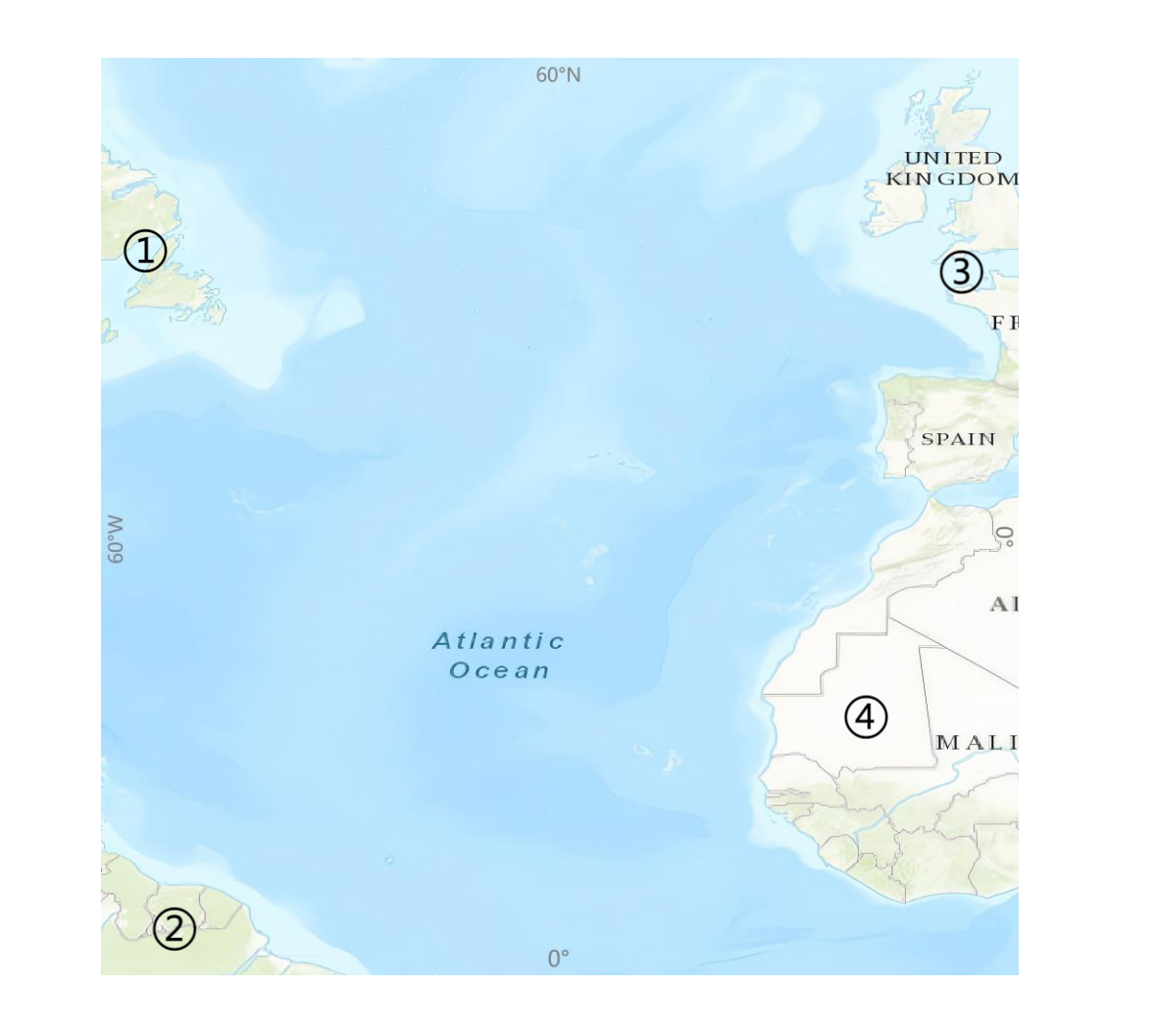}
        \caption{Map of the North Atlantic and its coastal regions} 
        \label{fig4.1(a)}
    \end{minipage}
    \hfill
    \begin{minipage}{0.47\textwidth}
        \centering
        \includegraphics[width=\textwidth]{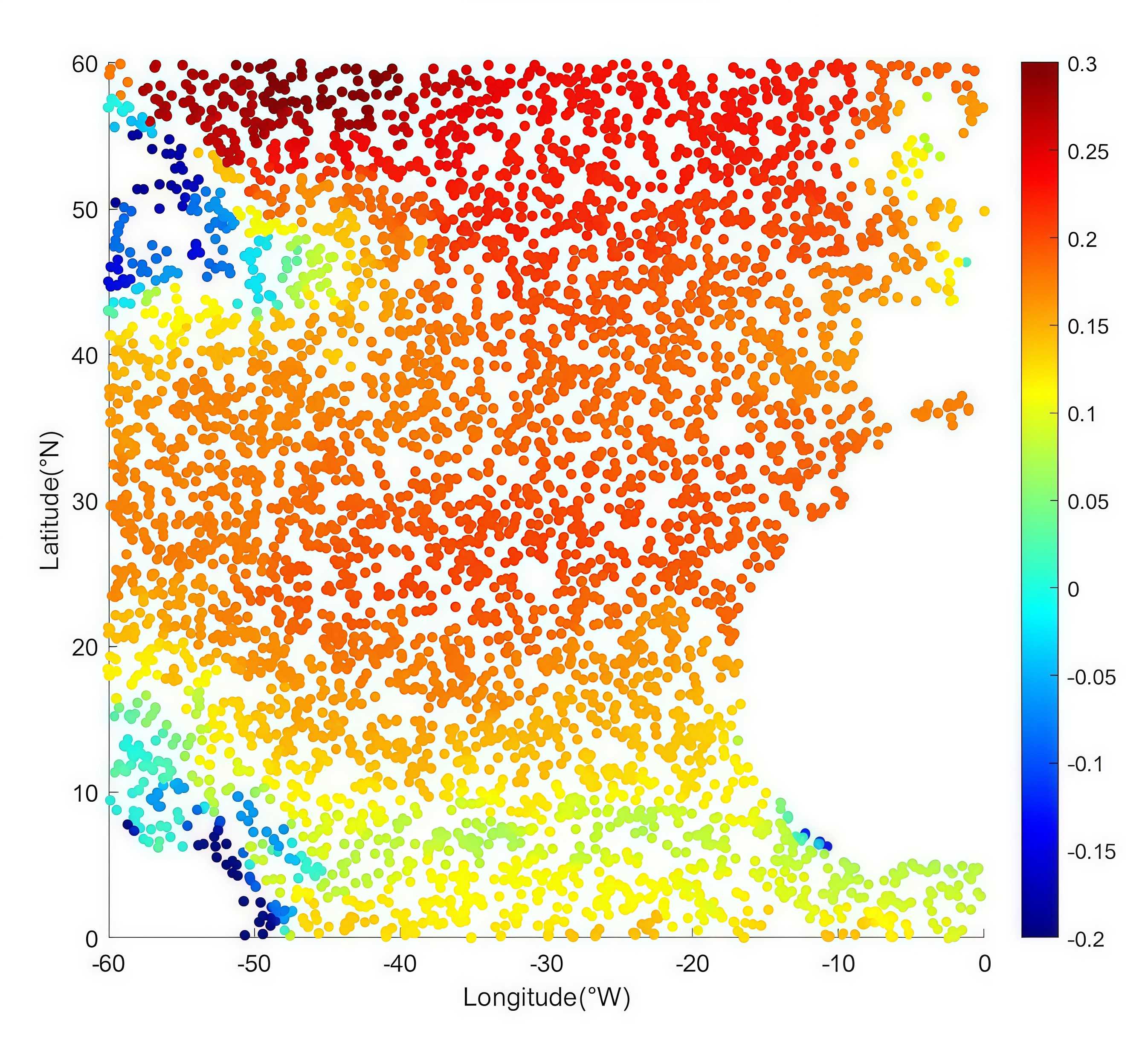}
        \caption{Estimated spatial coefficients for the T-S relationship based on the FLAT model}
        \label{fig4.1(b)}
    \end{minipage}
    \begin{minipage}{0.47\textwidth}
        \centering
        \includegraphics[width=\textwidth]{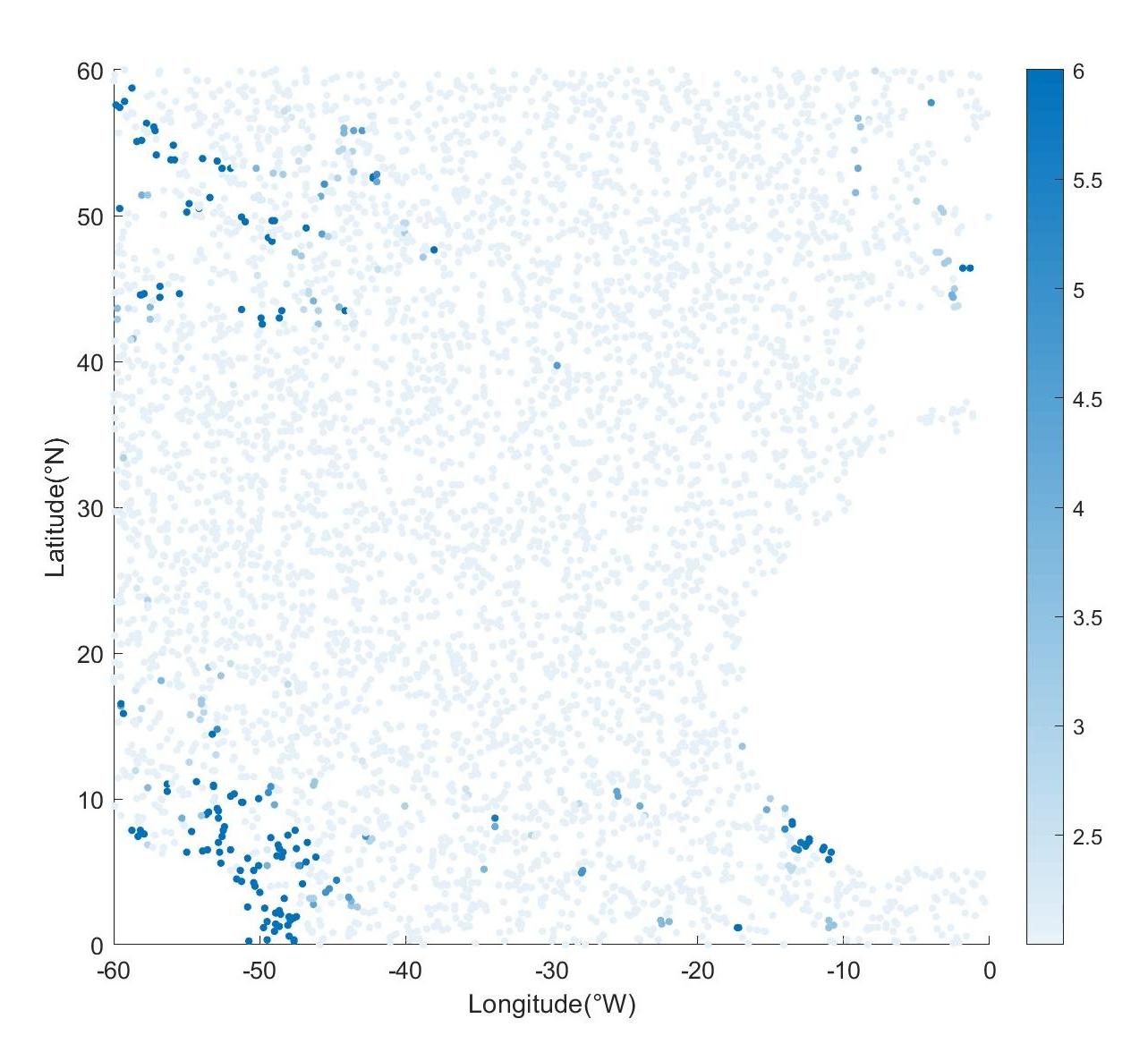}
        \caption{Estimated local SDQ in T-S coefficients}
        \label{fig4.1(c)}
    \end{minipage}
    \hfill
    \begin{minipage}{0.47\textwidth}
        \centering
        \includegraphics[width=\textwidth]{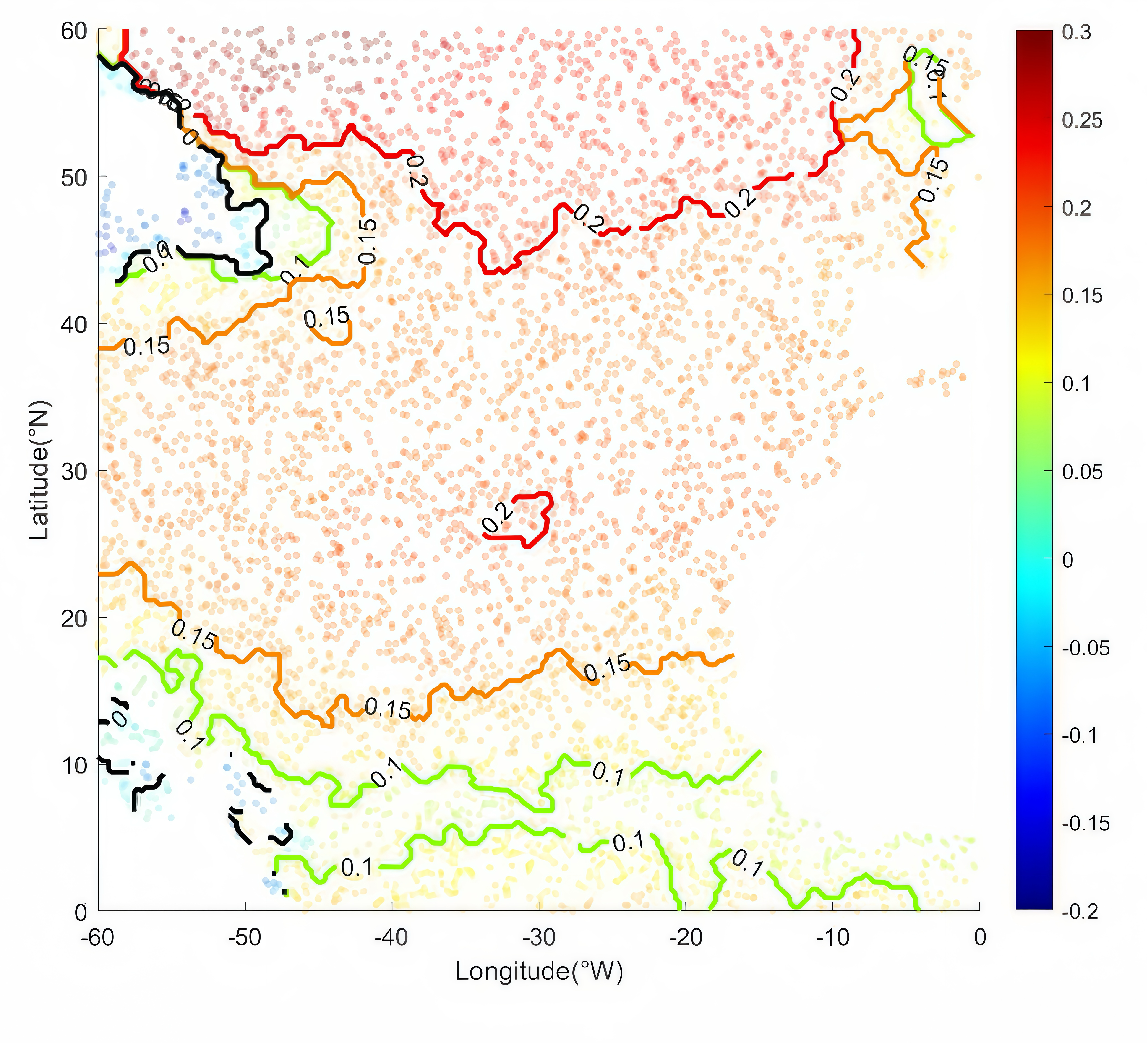}
        \caption{Identified T-S coefficient contours with clustering}
        \label{fig4.1(d)}
    \end{minipage}
\end{figure}

The real-world geographical environment of the corresponding area is depicted in Figure \ref{fig4.1(a)}. Based on the results estimated temperature-salinity relationship coefficients by FLAT in Figure \ref{fig4.1(b)}, coastal sea areas exhibit significant variations in T-S coefficients due to continental influences. For instance, regions near the eastern Atlantic coast of North America (the top-left region in Figure \ref{fig4.1(a)}), northeastern South America (the bottom-left region in Figure \ref{fig4.1(a)}), western Europe (the top-right region \textcircled{3} in Figure \ref{fig4.1(a)}), and northwestern Africa (the bottom-left region \textcircled{4} in Figure \ref{fig4.1(a)}), particularly estuarine zones, demonstrate pronounced changes in T-S relationships. This phenomenon may arise from localized processes such as evaporation, precipitation, and water mass mixing in coastal waters, including bays and straits, where temperature and salinity variations compensate for each other, thereby affecting seawater density structures. 

Additionally, in the open surface waters of the Atlantic Ocean, spanning from 60°N to the equator, T-S coefficients exhibit a pattern of initial decline, followed by a rise and subsequent decrease. This non-monotonic trend may reflect the combined effects of gyral circulation, large-scale current dynamics, and thermohaline processes. In polar-adjacent regions, low temperatures and seasonal freshwater fluxes contribute to distinct salinity patterns, in contrast to equatorial zones, where temperature increases are often accompanied by salinity shifts. These interacting factors jointly shape the spatial variability in T-S relationships. The findings enhance our understanding of thermohaline structures and their variability, which has important implications for the sustainable management of marine systems.


We further calculated the Spatial Difference Quotient (SDQ) of the sea surface T-S relationship based on the estimated coefficients from Figure \ref{fig4.1(b)}, as shown in Figure \ref{fig4.1(c)}. The SDQ can be interpreted as a numerical gradient of the T-S relationship at each spatial location; more details on its definition are provided in \ref{DRDA}. As shown in Figure \ref{fig4.1(c)}, FLAT effectively captures the spatial variation patterns of T-S relationships. Regions with larger difference quotients are primarily located near landmasses and the equator, particularly near two estuarine zones-the eastern Atlantic coast of North America and northeastern South America. In the equatorial zone, especially at 5°N latitude and between 20°E and 50°E longitude, the T-S coefficient approaches zero, while the SDQ remains relatively high. This pattern likely arises from the distinct marine conditions in equatorial waters: higher temperatures result in reduced density, while intense solar radiation drives strong evaporation and elevated salinity. The counteracting effects of temperature and salinity yield a near-zero T-S coefficient, and the sharp local contrasts contribute to the observed high spatial gradients, reflecting a dynamic T-S compensation regime.

In order to exploring the variation trend of the T-S relationship in the northern Atlantic Ocean more accurately, the spatial clustering of the T-S coefficient is conducted and the detected surface of discrete spatial heterogeneity is shown in Figure \ref{fig4.1(d)}. The T-S coefficient $\beta_1$ is consistently negative ($\beta_1 < 0$) in coastal estuarine regions, indicating a negative correlation between temperature and salinity in these zones. As one moves offshore into the open ocean, $\beta_1$ gradually increases and eventually becomes positive ($\beta_1 > 0$), reflecting a positive T-S correlation. Near the equator, contour lines where $\beta_1$ approaches zero emerge, forming what is referred to as the inverse temperature-salinity surface. This pattern highlights the spatial transition of T-S relationships across the region and suggests that temperature and salinity changes tend to be balanced in equatorial waters, with the coefficient converging to zero-indicative of a T-S compensation regime. Equatorial surface water, characterized by high temperatures and relatively low salinity, is influenced by intense year-round solar radiation that promotes strong evaporation. However, abundant precipitation delivers substantial freshwater input, offsetting salinity increases and giving rise to a T-S compensation effect, whereby changes in temperature and salinity are counterbalanced and $\beta_1$ approaches zero.


\subsection{Analysis of the T-S relationship in ocean cross-sections}

We also closely analyze the estimated T-S relationship in Atlantic Ocean cross-sections at a longitude of 25°W, spanning a latitude range from 60°N to 60°S, and a depth range from sea level to 3000 meters below sea level. 

\begin{figure}[H]
    \centering
    \begin{minipage}{0.9\textwidth}
        \centering
        \includegraphics[width=\textwidth]{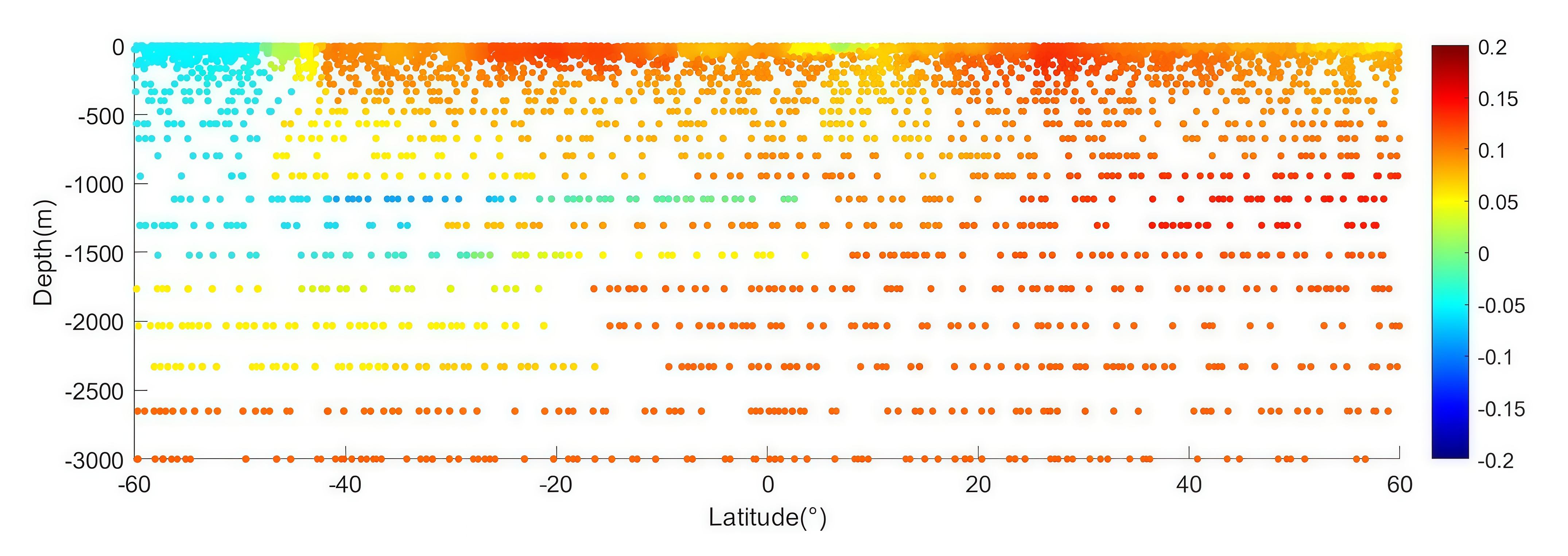}
        \caption{Vertical cross-section of the estimated T-S slope coefficients using FLAT}
        \label{fig4.2(a)}
    \end{minipage}
    \hfill
    \begin{minipage}{0.9\textwidth}
        \centering
        \includegraphics[width=\textwidth]{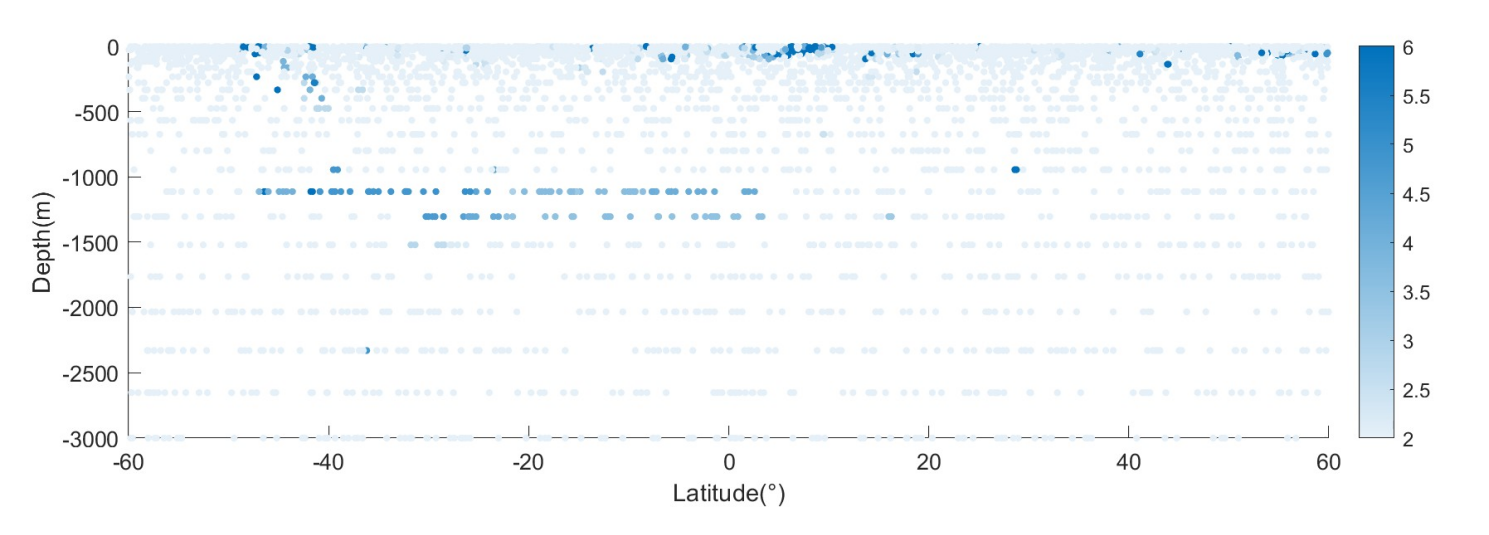}
        \caption{Vertical SDQ of the T-S coefficients along the ocean transect}
        \label{fig4.2(b)}
    \end{minipage}
    \begin{minipage}{0.9\textwidth}
        \centering
        \includegraphics[width=\textwidth]{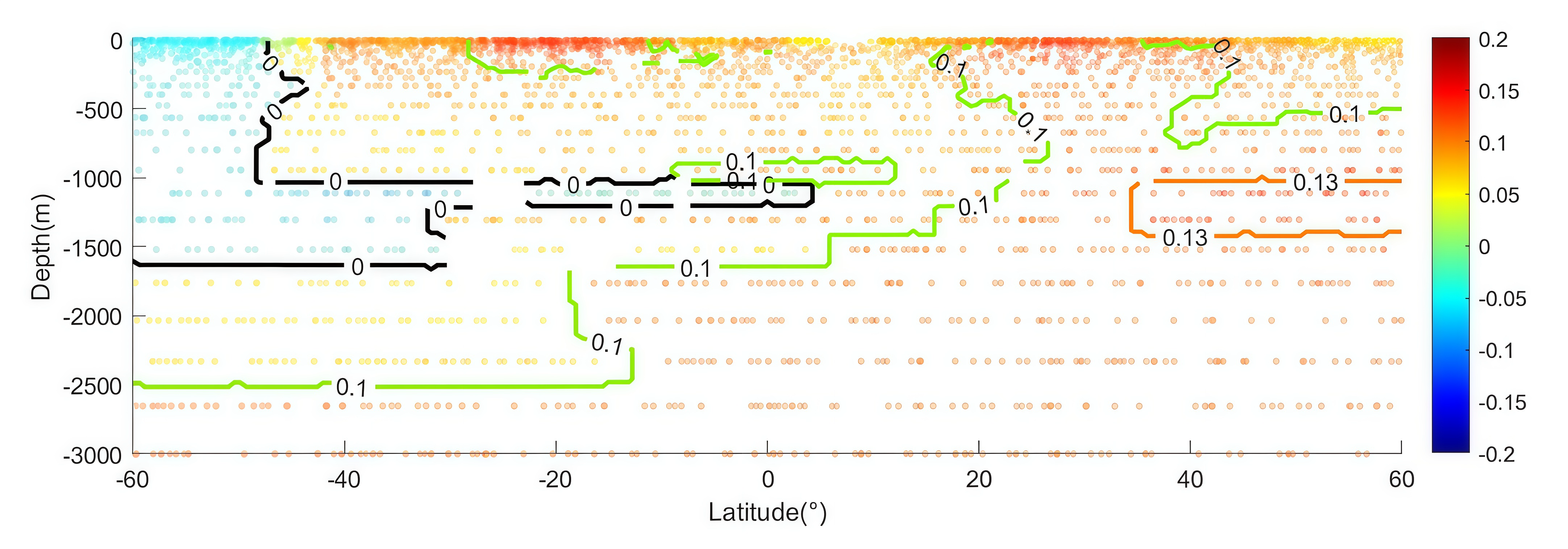}
        \caption{Identified T-S coefficient contours with clustering in the vertical cross-section}
        \label{fig4.2(c)}
    \end{minipage}
\end{figure}

Figure~\ref{fig4.2(a)} presents the estimation results of the T-S relationship coefficient $\beta_1$ for the ocean section obtained by FLAT. At a depth of approximately 1000 meters, spanning from 60°S latitude to near the equator, there is a notable abrupt change in the T-S relationship, with $\beta_1$ transitioning from negative to positive values. This pattern suggests a shift from a negative to a positive correlation between seawater temperature and salinity, which may be attributed to the physical characteristics and circulation patterns of distinct water masses. 

In the surface layers of Antarctic waters, temperature is extremely low while salinity is relatively high-a signature consistent with the thermohaline properties of Antarctic Intermediate Water. This is primarily influenced by the cold Antarctic climate, which leads to reduced seawater temperature. Meanwhile, the brine rejection process during sea ice formation increases salinity, thus resulting in a negative T-S correlation.

To better investigate the rate of change in the T-S relationship near the transition zone and more clearly identify thermohaline fronts, we calculated the SDQ of the T-S coefficient, as shown in Figure~\ref{fig4.2(b)}. This metric acts as a discrete analog to the spatial gradient and exhibits a sharp increase near front boundaries, indicating a rapid local change in T-S structure.

To further reveal the spatial patterns of T-S characteristics, we employ the FLAT method to first estimate local coefficients and then conduct clustering to obtain discrete regions with similar thermohaline behavior. Figure~\ref{fig4.2(c)} displays the contour plot of T-S coefficients across depth and latitude. The results show that the FLAT model successfully delineates T-S regimes in deep ocean layers, capturing the thermohaline features of water masses. Notably, $\beta_1$ tends to increase with depth, suggesting that salinity variations contribute more significantly to density changes at greater depths than temperature does.

When $\beta_1<0$, temperature and salinity exhibit a negative correlation. Consequently, as seawater temperature decreases, its density increases, simultaneously, a rise in salinity also leads to an increase in density, making the water mass denser and causing the seawater to sink, forming a specific water mass. This process has a significant impact on the global thermohaline circulation. Particularly in deep ocean waters, this phenomenon of density increase facilitates vertical mixing and heat transfer within the ocean. Conversely, when $\beta_1>0$, temperature and salinity display a positive correlation. In this scenario, changes in seawater density are influenced by various factors including temperature gradients, salinity gradients, thermal expansion coefficients, and haline contraction coefficients, rendering the situation more complex. Among these factors, if the effects of temperature and salinity on seawater density offset each other, it results in thermohaline compensation.

Each water mass possesses specific temperature and salinity ranges, along with a unique thermohaline relationship that arises from these parameters. For instance, Antarctic Intermediate Water is characterized by low temperature and high salinity, whereas Equatorial Surface Water exhibits high temperature and low salinity. By analyzing temperature and salinity data of water bodies, water mass characteristics can be identified. Furthermore, the thermohaline relationship can also reveal the evolution of water masses. As water masses move and disperse in the ocean, both temperature and salinity undergo changes, reflecting the interaction and evolution process between the water mass and its surrounding environment. Consequently, by analyzing thermohaline data from different locations and times, it is possible to trace the movement trajectory and evolutionary trends of water masses.

Spatial clustering analysis facilitates a deeper understanding of the variation patterns of the T-S relationship in different sea areas of the Atlantic Ocean and holds significance for marine climate change studies. For instance, water masses such as Antarctic Intermediate Water and equatorial surface water are vital components of the global ocean circulation. Their T-S compensation phenomena influence the operation of the global thermohaline circulation. Warm, low-salinity seawater in equatorial regions is transported to high-latitude areas via ocean currents, where it cools and increases in salinity before sinking to form deep seawater. This process participates in the global oceanic conveyor belt and subsequently exerts significant impacts on global climate.

\section{Conclusion and Discussion}  
\label{S5}

We proposed a new approach for modeling spatial heterogeneity, named the Fused Lasso regression model with an Adaptive minimum spanning Tree (FLAT). FLAT is capable of estimating continuous spatial heterogeneity in spatial regression models and simultaneously selecting relevant high-dimensional features, while preserving data-driven detection of discrete spatial heterogeneity through adaptive fused lasso regularization terms. Comprehensive simulation studies demonstrate the excellent finite-sample performance of FLAT. Moreover, the application to oceanographic data on temperature-salinity (T-S) relationships further illustrates the practical usefulness of the proposed method in empirical analyses. While the proposed FLAT framework effectively reconciles continuous and discrete spatial heterogeneity in regression modeling, several potential extensions can be valuable further topics, such as incorporating temporal dynamics into the adaptive fused regularization framework to model spatio-temporal heterogeneity and including uncertainty quantification to improve the interpretability and robustness of the algorithms.

\bibliographystyle{abbrvnat}
\bibliography{reference}

\begin{thebibliography}{34}
\providecommand{\natexlab}[1]{#1}
\providecommand{\url}[1]{\texttt{#1}}
\expandafter\ifx\csname urlstyle\endcsname\relax
  \providecommand{\doi}[1]{doi: #1}\else
  \providecommand{\doi}{doi: \begingroup \urlstyle{rm}\Url}\fi

\bibitem[Anselin(2010)]{anselin2010thirty}
L.~Anselin.
\newblock Thirty years of spatial econometrics.
\newblock \emph{Papers in Regional Science}, 89\penalty0 (1):\penalty0 3--26, 2010.

\bibitem[Anselin(2013)]{anselin2013spatial}
L.~Anselin.
\newblock \emph{Spatial econometrics: methods and models}, volume~4.
\newblock Springer Science \& Business Media, 2013.

\bibitem[Bailey et~al.(1995)Bailey, Gatrell, et~al.]{bailey1995interactive}
T.~C. Bailey, A.~C. Gatrell, et~al.
\newblock \emph{Interactive spatial data analysis}, volume 413.
\newblock Longman Scientific \& Technical Essex, 1995.

\bibitem[Brunsdon et~al.(1996)Brunsdon, Fotheringham, and Charlton]{brunsdon1996geographically}
C.~Brunsdon, A.~S. Fotheringham, and M.~E. Charlton.
\newblock Geographically weighted regression: a method for exploring spatial nonstationarity.
\newblock \emph{Geographical Analysis}, 28\penalty0 (4):\penalty0 281--298, 1996.

\bibitem[Chen et~al.(2024)Chen, Zhang, Ma, and Fang]{chen2024heterogeneity}
Y.~Chen, Q.~Zhang, S.~Ma, and K.~Fang.
\newblock Heterogeneity-aware clustered distributed learning for multi-source data analysis.
\newblock \emph{Journal of Machine Learning Research}, 25\penalty0 (211):\penalty0 1--60, 2024.

\bibitem[Duque et~al.(2011)Duque, Church, and Middleton]{duque2011p}
J.~C. Duque, R.~L. Church, and R.~S. Middleton.
\newblock The p-regions problem.
\newblock \emph{Geographical Analysis}, 43\penalty0 (1):\penalty0 104--126, 2011.

\bibitem[Ester et~al.(1996)Ester, Kriegel, Sander, and Xu]{ester1996density}
M.~Ester, H.-P. Kriegel, J.~Sander, and X.~Xu.
\newblock A density-based algorithm for discovering clusters in large spatial databases with noise.
\newblock volume~96, pages 226--231, 1996.

\bibitem[Ferrari et~al.(2001)Ferrari, Paparella, Rudnick, and Young]{ferrari2001temperature}
R.~Ferrari, F.~Paparella, D.~Rudnick, and W.~Young.
\newblock The temperature-salinity relationship of the mixed layer.
\newblock \emph{From Stirring to Mixing in a Stratified Ocean}, pages 95--104, 2001.

\bibitem[Flament et~al.(1985)Flament, Armi, and Washburn]{flament1985evolving}
P.~Flament, L.~Armi, and L.~Washburn.
\newblock The evolving structure of an upwelling filament.
\newblock \emph{Journal of Geophysical Research: Oceans}, 90\penalty0 (C6):\penalty0 11765--11778, 1985.

\bibitem[Fotheringham et~al.(2017)Fotheringham, Yang, and Kang]{fotheringham2017multiscale}
A.~S. Fotheringham, W.~Yang, and W.~Kang.
\newblock Multiscale geographically weighted regression (mgwr).
\newblock \emph{Annals of the American Association of Geographers}, 107\penalty0 (6):\penalty0 1247--1265, 2017.

\bibitem[Gelfand et~al.(2003)Gelfand, Kim, Sirmans, and Banerjee]{GelfandJASA}
A.~E. Gelfand, H.-J. Kim, C.~F. Sirmans, and S.~Banerjee.
\newblock Spatial modeling with spatially varying coefficient processes.
\newblock \emph{Journal of the American Statistical Association}, 98\penalty0 (462):\penalty0 387--396, 2003.
\newblock \doi{10.1198/016214503000170}.
\newblock URL \url{https://doi.org/10.1198/016214503000170}.

\bibitem[Kudela et~al.(2015)Kudela, Palacios, Austerberry, Accorsi, Guild, and Torres-Perez]{kudela2015application}
R.~M. Kudela, S.~L. Palacios, D.~C. Austerberry, E.~K. Accorsi, L.~S. Guild, and J.~Torres-Perez.
\newblock Application of hyperspectral remote sensing to cyanobacterial blooms in inland waters.
\newblock \emph{Remote Sensing of Environment}, 167:\penalty0 196--205, 2015.

\bibitem[Lam and Souza(2020)]{lam2020estimation}
C.~Lam and P.~C. Souza.
\newblock Estimation and selection of spatial weight matrix in a spatial lag model.
\newblock \emph{Journal of Business \& Economic Statistics}, 38\penalty0 (3):\penalty0 693--710, 2020.

\bibitem[LeSage and Pace(2009)]{lesage2009introduction}
J.~LeSage and R.~K. Pace.
\newblock \emph{Introduction to spatial econometrics}.
\newblock Chapman and Hall/CRC, 2009.

\bibitem[Li and Sang(2019)]{liSangJASA2019spatial}
F.~Li and H.~Sang.
\newblock Spatial homogeneity pursuit of regression coefficients for large datasets.
\newblock \emph{Journal of the American Statistical Association}, 114\penalty0 (527):\penalty0 1050--1062, 2019.

\bibitem[Li and Yang(2021)]{li2021thermohaline}
J.~Li and Y.~Yang.
\newblock Thermohaline interleaving induced by horizontal temperature and salinity gradients from above.
\newblock \emph{Journal of Fluid Mechanics}, 927:\penalty0 A12, 2021.

\bibitem[Lin et~al.(2023)Lin, Xu, Liu, Hu, Zhang, and Cao]{lin2023scale}
H.~Lin, S.~Xu, Z.~Liu, J.~Hu, F.~Zhang, and Z.~Cao.
\newblock Scale-dependent temperature-salinity compensation in frontal regions of the taiwan strait.
\newblock \emph{Journal of Geophysical Research: Oceans}, 128\penalty0 (2):\penalty0 e2022JC019134, 2023.

\bibitem[Luo et~al.(2021)Luo, Sang, and Mallick]{luo2021bayesian}
Z.~T. Luo, H.~Sang, and B.~Mallick.
\newblock A bayesian contiguous partitioning method for learning clustered latent variables.
\newblock \emph{Journal of Machine Learning Research}, 22\penalty0 (37):\penalty0 1--52, 2021.

\bibitem[Miller and Han(2009)]{miller2009geographic}
H.~J. Miller and J.~Han.
\newblock \emph{Geographic data mining and knowledge discovery}.
\newblock CRC press, 2009.

\bibitem[Rahmstorf(2003)]{rahmstorf2003thermohaline}
S.~Rahmstorf.
\newblock Thermohaline circulation: The current climate.
\newblock \emph{Nature}, 421\penalty0 (6924):\penalty0 699--699, 2003.

\bibitem[Raudenbush(2002)]{raudenbush2002hierarchical}
S.~W. Raudenbush.
\newblock Hierarchical linear models: Applications and data analysis methods.
\newblock \emph{Advanced Quantitative Techniques in the Social Sciences Series/SAGE}, 2002.

\bibitem[Reid(1969)]{reid1969sea}
J.~Reid.
\newblock Sea-surface temperature, salinity, and density of pacific ocean in summer and in winter.
\newblock \emph{Deep-Sea Research}, page 215, 1969.

\bibitem[Roden(1975)]{roden1975north}
G.~I. Roden.
\newblock On north pacific temperature, salinity, sound velocity and density fronts and their relation to the wind and energy flux fields.
\newblock \emph{Journal of Physical Oceanography}, 5\penalty0 (4):\penalty0 557--571, 1975.

\bibitem[Rudnick and Luyten(1996)]{rudnick1996intensive}
D.~L. Rudnick and J.~R. Luyten.
\newblock Intensive surveys of the azores front: 1. tracers and dynamics.
\newblock \emph{Journal of Geophysical Research: Oceans}, 101\penalty0 (C1):\penalty0 923--939, 1996.

\bibitem[Talley(2011)]{talley2011descriptive}
L.~D. Talley.
\newblock \emph{Descriptive physical oceanography: an introduction}.
\newblock Academic press, 2011.

\bibitem[Talley(2013)]{talley2013closure}
L.~D. Talley.
\newblock Closure of the global overturning circulation through the indian, pacific, and southern oceans: Schematics and transports.
\newblock \emph{Oceanography}, 26\penalty0 (1):\penalty0 80--97, 2013.

\bibitem[Tibshirani(1996)]{tibshirani1996regression}
R.~Tibshirani.
\newblock Regression shrinkage and selection via the lasso.
\newblock \emph{Journal of the Royal Statistical Society Series B: Statistical Methodology}, 58\penalty0 (1):\penalty0 267--288, 1996.

\bibitem[Van~Dusen and Nissen(2019)]{van2019modernizing}
B.~Van~Dusen and J.~Nissen.
\newblock Modernizing use of regression models in physics education research: A review of hierarchical linear modeling.
\newblock \emph{Physical Review Physics Education Research}, 15\penalty0 (2):\penalty0 020108, 2019.

\bibitem[Velasco et~al.(2021)Velasco, Estrada, Calder{\'o}n-Bustamante, Swingedouw, Ureta, Gay, and Defrance]{velasco2021synergistic}
J.~A. Velasco, F.~Estrada, O.~Calder{\'o}n-Bustamante, D.~Swingedouw, C.~Ureta, C.~Gay, and D.~Defrance.
\newblock Synergistic impacts of global warming and thermohaline circulation collapse on amphibians.
\newblock \emph{Communications biology}, 4\penalty0 (1):\penalty0 141, 2021.

\bibitem[Ward and Gleditsch(2018)]{ward2018spatial}
M.~D. Ward and K.~S. Gleditsch.
\newblock \emph{Spatial regression models}, volume 155.
\newblock Sage Publications, 2018.

\bibitem[Xing et~al.(2024)Xing, Wan, Wen, and Zhong]{xing2024golfs}
Z.~Xing, Y.~Wan, J.~Wen, and W.~Zhong.
\newblock Golfs: feature selection via combining both global and local information for high dimensional clustering.
\newblock \emph{Computational Statistics}, 39\penalty0 (5):\penalty0 2651--2675, 2024.

\bibitem[Xing et~al.(2025)Xing, Tan, Zhong, and Shi]{xing2025calms}
Z.~Xing, H.~Tan, W.~Zhong, and L.~Shi.
\newblock Calms: Constrained adaptive lasso with multi-directional signals for latent networks reconstruction.
\newblock \emph{Neurocomputing}, 630:\penalty0 129545, 2025.

\bibitem[Zhang et~al.(2024)Zhang, Liu, and Zhu]{zhang2024learning}
X.~Zhang, J.~Liu, and Z.~Zhu.
\newblock Learning coefficient heterogeneity over networks: A distributed spanning-tree-based fused-lasso regression.
\newblock \emph{Journal of the American Statistical Association}, 119\penalty0 (545):\penalty0 485--497, 2024.

\bibitem[Zou(2006)]{zou2006adaptive}
H.~Zou.
\newblock The adaptive lasso and its oracle properties.
\newblock \emph{Journal of the American Statistical Association}, 101\penalty0 (476):\penalty0 1418--1429, 2006.

\end{thebibliography}

\newpage

\section*{Appendix}  

\subsection{FLAT algorithm details}
\label{FLATdetailsalgo}
As introduced in Section \ref{S2}, we set the adaptive weights as the spatial distance between corresponding spatial locations, then the regularization terms based on adaptive-minimum-spanning tree can be further formulated as 
\begin{align}
\label{xll}
    P_{\lambda_1}(\boldsymbol{\beta}) 
    &= \sum_{(s_i, s_j) \in \mathbb{E}} P_{\lambda_1} \left( \boldsymbol{\beta}(s_i) - \boldsymbol{\beta}(s_j) \right)  \nonumber \\
     &= \lambda_1 \sum_{k=1}^{p} \sum_{i=1}^{n} \sum_{j \in \mathcal{N}_i} \pi_{(s_i,s_j)} \left| \beta_k(s_i) - \beta_k(s_j) \right|,
\end{align}
where $\mathbb{E}$ represents the edge set derived from the minimum spanning tree, and $\mathcal{N}_i$ denotes the neighboring nodes of node $i$ in the minimum spanning tree. With some derivation, the regularization for spatial heterogeneity can be rewritten as
\begin{equation}
    P_{\lambda_1}(\boldsymbol{\beta}) = \lambda_1 \sum_{k=1}^{p} \| \bm{\pi} \bm{H} \bm{\beta}_k \|_1,
\end{equation}
based on the definition of matrix $\bm{H}$, where the sparse weight matrix $\bm{\pi} \in \mathbb{R}^{(n-1) \times (n-1)}$ contains all the weights of $n-1$ potential edges. For simplicity, we use matrix notation and rewrite the objective function of FLAT \eqref{FLAT_original} as
\begin{equation}
\label{MLFLAT}
    L_{\text{FLAT}}(\boldsymbol{\beta}) = \frac{1}{2} \| \mathbf{y} - \bm{x} \boldsymbol{\beta} \|_2^2 + \lambda_1 \sum_{k=1}^{p} \|\bm{\pi} \bm{H} \bm{\beta}_k \|_1 + \lambda_2 \sum_{k=1}^{p} \|\bm{\beta}_k \|_1,
\end{equation}
where \(\mathbf{y} = [y(s_1), ..., y(s_n)]^\top \in \mathbb{R}^{n}\) represents the vector of response values for $n$ locations. \( \bm{x} \triangleq [  \text{diag}(\bm{x}_1), ..., \text{diag}(\bm{x}_{p})] \in \mathbb{R}^{n \times np} \), where \( \bm{x}_k \) represents the \( k \)-th covariate vector of length \( n \), and the observation vector is \( \bm{x}_k \triangleq (x_k(s_1), ..., x_k(s_n))^\top \in \mathbb{R}^{n} \) is the corresponding design matrix with spatial feature. We denote \( \boldsymbol{\beta} = [\boldsymbol{\beta}^\top(s_1), ..., \boldsymbol{\beta}^\top (s_n)]^\top \in \mathbb{R}^{np} \), where \( \boldsymbol{\beta}(s_i) \triangleq [\beta_1(s_i), ..., \beta_p(s_i)]^\top \in \mathbb{R}^{p} \).           
Then we define the matrix $\tilde{\bm{H}}$ and reparameterize coefficient matrix accordingly 
\begin{equation}
\label{repara_Htilde}
    \tilde{\bm{H}} \triangleq \begin{bmatrix} \pi \bm{H} \\ \frac{\lambda_2}{\lambda_1} \bm{I} \end{bmatrix} \in \mathbb{R}^{n \times n}, ~~ \boldsymbol{\theta}_k = \begin{bmatrix} \pi \bm{H} \\ \frac{\lambda_2}{\lambda_1} \bm{I} \end{bmatrix} \boldsymbol{\beta}_k = \tilde{\bm{H}} \boldsymbol{\beta}_k,
\end{equation}
where \( \boldsymbol{\theta}_k \triangleq [\theta_k(s_1), ..., \theta_k(s_n)]^\top \in \mathbb{R}^{n} \) is the parameter vector and $\bm{I}$ denotes the . 
Since \( \tilde{\bm{H}} \) is invertible, we reparameterize $\bm{\beta}$ as equation \eqref{repara_Htilde} and express \( \boldsymbol{\beta}_k \) as \( \boldsymbol{\beta}_k = \tilde{\bm{H}}^{-1} \boldsymbol{\theta}_k \). Based on this transformation, we reparameterize the design matrix as 
\begin{equation}
\tilde{\bm{X}} \triangleq \left[\text{diag}(\bm{x}_1) \tilde{\bm{H}}^{-1}, ..., \text{diag}(\bm{x}_p) \tilde{\bm{H}}^{-1}\right].
\end{equation}

With the reparameterization defined above, we rewrite the object function of proposed FLAT approach in \eqref{FLAT_original} as       
\begin{equation} 
\label{ReparameterizationFLAT}
L_{\text{FLAT}}(\boldsymbol{\theta}) = \frac{1}{2} \|\mathbf{y} - \tilde{\bm{X}}\boldsymbol{\theta} \|_2^2 + \lambda_1 \sum_{k=1}^{p} \|\boldsymbol{\theta}_k \|_1,
\end{equation}
where \( \boldsymbol{\theta} \triangleq [\boldsymbol{\theta}_1^\top, ..., \boldsymbol{\theta}_p^\top]^\top \in \mathbb{R}^{np \times 1} \). We define the FLAT estimator of the coefficients as $\hat{\boldsymbol{\theta} }_{\mathrm{FLAT}} = \arg\min_{\boldsymbol{\theta} \in \Theta} L_{\text{FLAT}}(\boldsymbol{\theta})$.  We can obtain the FLAT estimator of original regression coefficients using \( \hat{\boldsymbol{\beta}}_{\mathrm{FLAT}} = \tilde{\bm{H}}^{-1} \hat{\boldsymbol{\theta}}_{\mathrm{FLAT}} \). 

By minimizing \eqref{ReparameterizationFLAT}, $\hat{\boldsymbol{\theta} }_{\mathrm{FLAT}}$ can detect the discrete spatial heterogeneity in effects while estimating the continuous spatial heterogeneity, and also select high-dimensional spatial features. As $L_1$ penalty terms are not differentiable at the point $0$, the standard soft-thresholding can be used to derive the updates in optimization algorithm \citep{ tibshirani1996regression, xing2025calms}. To indicate non-convex spatial clusters in terms of feature effects, some classic density-based clustering algorithm \citep{ester1996density, miller2009geographic} can be used to further obtain the spatial surface in specific applications, such as the inverse temperature-salinity surface in thermohaline circulation.

\begin{algorithm}[h]
\caption{FLAT Algorithm with Adaptive-minimum-spanning Tree}
\label{alg:FLAT}
\begin{tabular}{l l} 
1: & \textbf{Input:} Spatial data $\{(\bm{x}(s_i), y(s_i))\}_{i=1}^n$, parameters $\lambda_1, \lambda_2, \gamma$ \\
2: & Initialization of pairwise distances via \eqref{eq:beta_distance} and obtain the adaptive weights $\pi_{(s_i,s_j)}$ in \eqref{xll} \\
3: & Obtain the fused structure $\bm{H}$ in \eqref{Hbeta1} based on edge set $\mathbb{E}$, which is built by MST \\
   & from Prim's algorithm \\
4: & Reparameterization defined in \eqref{repara_Htilde} Compute adaptive weights $\bm{\pi}$ in \eqref{MLFLAT} \\
5: & Formulate objective function and conduct reparameterization via \eqref{ReparameterizationFLAT} \\
6: & \textbf{repeat} \\
7: & \quad \textbf{for} $j = 1$ to $np$ \textbf{do} \\
8: & \quad \quad Update $
\theta_j^{(t+1)} = \mathcal{S}_{\lambda_1} \{\theta_j^{(t)} - \eta \sum_{i=1}^n \tilde{X}_{ij}(y_i - \hat{y}_i^{(t)} + \tilde{X}_{ij}\theta_j^{(t)}) \}
$ \\
   & \quad \quad  where $\mathcal{S}_{\lambda}(x) = \text{sign}(x)(|x| - \lambda)_+$ is soft-thresholding operator 
\\
9: & \quad \textbf{end for} \\
10:& \textbf{until} $\|\boldsymbol{\theta}^{(t+1)} - \boldsymbol{\theta}^{(t)}\|_2 < 10^{-6}$ \\
11:& Obtain $\hat{\boldsymbol{\theta}}_{\mathrm{FLAT}}$ and the estimated coefficients $\hat{\boldsymbol{\beta}}_{\mathrm{FLAT}}$ \\
12:& Initialization of the hyper-parameter sequences $\epsilon$ and $minPts$, set $cluster.id=0$ \\
13:& \textbf{for} each $(\epsilon, minPts)$ \textbf{do} \\
14:& \quad Initialize all locations as \textit{unvisited} \\
15:& \quad \textbf{for} each \textit{unvisited} location $s_i$ \textbf{do} \\
16:& \quad \quad Obtain its $\epsilon$-neighborhood $N_\epsilon(s_i) = \{s_j : \|\hat{\bm{\beta}}_{\mathrm{FLAT}}(s_j) - \hat{\bm{\beta}}_{\mathrm{FLAT}}(s_i)\|_2 \leq \epsilon \}$ \\
17:& \quad \quad \textbf{if} $|N_\epsilon(s_i)| \geq minPts$ \textbf{then} \\
18:& \quad \quad \quad Create new cluster: $cluster.id \gets cluster.id + 1$ \\
19:& \quad \quad \quad \textbf{for} each $s_j \in N_\epsilon(s_i)$ \textbf{do} \\
20:& \quad \quad \quad \quad Compute $N_\epsilon(s_j)$ \\
21:& \quad \quad \quad \quad \textbf{if} $|N_\epsilon(s_j)| \geq minPts$ \textbf{then} \\
22:& \quad \quad \quad \quad \quad Merge $N_\epsilon(s_j)$ into $N_\epsilon(s_i)$ \\
23:& \quad \quad \quad \quad \textbf{else} \\
24:& \quad \quad \quad \quad \quad Mark $s_j$ as \textit{noise} \\
25:& \quad \quad \quad \quad \textbf{end if} \\
26:& \quad \quad \quad \textbf{end for} \\
27:& \quad \quad \quad Expand cluster: Add $N_\epsilon(s_i)$ to cluster $C_{cluster.id}$ \\
28:& \quad \quad \quad Mark all locations in $C_{cluster.id}$ as \textit{visited} \\
29:& \quad \quad \textbf{end if} \\
30:& \quad \textbf{end for} \\
31:& \quad Evaluate and select the clustering labels \\
32:& \textbf{end for} \\
33:& \textbf{Output:} Estimated coefficients $\hat{\boldsymbol{\beta}}_{\mathrm{FLAT}}$, detected spatial clusters \\ 
\end{tabular}
\end{algorithm}

\subsection{The evaluations of clustering}
\label{DEC}

Many evaluations are widely used to measure the results of clustering, such as the rand Index and adjusted Rand index with known clustering labels, and Silhouette coefficient and Calinski-Harabasz index without true labels. In this section, we will introduce the these measures and more discussion and usage can be find in \citet{xing2024golfs}.

The Rand Index is a crucial metric used to measure the similarity between clustering results and true class labels. If $a$ represents the number of pairs of samples that belong to the same class in both the clustering result and the true class labels, and $b$ represents the number of pairs of samples that belong to different classes in both, the formula for calculating RI is written as
\begin{equation}
    \text{RI}=\frac{a+b}{\binom{n}{2}}.
\end{equation}

Since the RI does not yield a value of zero for random clustering results, the Adjusted Rand Index (ARI) is frequently employed. It corrects for the influence of random clustering, with a value range from -1 to 1, where an ARI of 0 indicates purely random clustering. The ARI addresses the limitation of RI, which tends to exhibit spuriously high values when the number of clusters is large due to chance agreements. Additionally, it enables direct comparison of results across different clustering algorithms. The expected value of RI under random clustering is denoted as $\text{E}(\text{RI})$. Then, the ARI is computed as
\begin{equation}
    \text{ARI}=\frac{\text{RI}-\text{E}(\text{RI})}{\text{max}(\text{RI})-\text{E}(\text{RI})}.
\end{equation}

The Silhouette Coefficient (SC) measures clustering effectiveness by evaluating the clustering quality of individual sample points within a cluster. This coefficient comprehensively considers the internal consistency and external separation of sample clusters. The SC ranges from -1 to 1, where a value closer to 1 indicates better clustering performance, a value close to -1 suggests that the sample point is misclustered, and a value near 0 implies uncertainty in determining which cluster a sample belongs to. Assuming there is a sample point $s_i$, the average distance between this sample point and samples within the same cluster is denoted as $a(s_i)$, and the average distance to samples from other clusters is denoted as $b(s_i)$. Then, the SC for sample point $s_i$ is calculated as
\begin{equation}
    \text{SC}(s_i)=\frac{b(s_i)-a(s_i)}{\max\{a(s_i),b(s_i)\}}.
\end{equation}
Accordingly, SC for clustering is defined as the average of the SC for all sample points
\begin{equation}
    \text{SC}=\frac{1}{n}\sum\limits_{i=1}^{n}\text{SC}(s_i).
\end{equation}

Calinski-Harabasz Index (CHI), also called Variance Ratio Criterion (VRC), measures clustering effectiveness using the ratio of the distance between groups and the distance within group. Larger value of CHI indicates better result of clustering. Assuming there are $k$ clusters in the clustering of variable $\bm{x}$, with $n_i$ samples in each cluster, we note the centroid of cluster $C_i$ as $\mathbf{c_i}$ and the overall centroid of the data as $\mathbf{c}$. The CHI can be expressed as
\begin{equation}
\text{CHI}=\frac{\sum\limits_{i=1}^{k}n_i\|\mathbf{c}_i-\mathbf{c}\|^2/(k-1)}{\sum\limits_{i=1}^{k}\sum\limits_{\bm{x}\in C_i}\|\bm{x}-\mathbf{c}_i\|^2/(n-k)}.
\end{equation}

\subsection{Details of the real-data analysis}
\label{DRDA}
Taking two-dimensional space as an example, suppose an observation point $s_i$ , whose gradient points projected onto the horizontal and vertical axes are denoted as $s_{i_1}$ and $s_{i_2}$, respectively. If the observed value at each point $s$ in space is represented by $\beta(s)$, then the Spatial Difference Quotient (SDQ) at point $s_i$ can be expressed as:
\begin{equation}
    D(s_i)=\sqrt{\frac{\left(\beta(s_i)-\beta(s_{i_1})\right)^2}{d_1^2}+\frac{\left(\beta(s_i)-\beta(s_{i_2})\right)^2}{d_2^2}},
\end{equation}
Here, $d_1$ and $d_2$ denote the Euclidean distances from point $s_i$ to $s_{i_1}$ and $s_{i_2}$, respectively. Since the observed values at the two gradient points in the aforementioned formula may not exist, we instead consider the two nearest observation points to $s_i$, denoted as $s_j$ and $s_k$. Consequently, $D(s_i)$ is defined as:
\begin{equation}
    D(s_i)=\sqrt{\frac{\left(\beta(s_i)-\beta(s_j)\right)^2}{d_{ij}^2\sin^2\gamma}+\frac{\left(\beta(s_i)-\beta(s_k)\right)^2}{d_{ik}^2\sin^2\gamma}-2\frac{\left(\beta(s_i)-\beta(s_j)\right)\left(\beta(s_i)-\beta(s_k)\right)\cos\gamma}{d_{ij}d_{ik}\sin^2\gamma}},
\end{equation}
Here, $\gamma$ denotes the included angle between vector $\overrightarrow{s_is_j}$ and vector $\overrightarrow{s_is_k}$.



\end{document}